\documentclass{elsart}

\usepackage{amsmath,epsfig}
\usepackage{amssymb}

\def\be{\begin{equation}}
\def\ee{\end{equation}}
\def\bmlt{\begin{multline}}
\def\emlt{\end{multline}}

\begin{document}

\begin{frontmatter}

\title{
Correlations in impact-parameter
space\\
in a hierarchical saturation model\\
for QCD at high energy
}

\author{A. H. Mueller}
\address{Department of physics, Columbia University, New York, USA}
\author{S. Munier}
\address{Centre de physique th{\'e}orique, \'Ecole Polytechnique,
CNRS, Palaiseau, France}

\date{\today}

\begin{abstract}
In order to get an estimate of the homogeneity 
of the distribution of matter in a fast hadron or nucleus,
we compute the 
correlations of the saturation scales $Q_s$ between
different points
in impact-parameter space,
in some specific saturation models.
We find that these correlations are quite strong:
The saturation scale is nearly uniform in domains
whose sizes scale like 
$\exp\left[\text{const}\times\ln^2(1/\alpha_s^2)\right]/Q_s$, which
means that the density of gluons should not
fluctuate significantly
over regions of that typical size.
We expect these conclusions as well as
the explicit analytical expressions we obtain 
for the correlations
to be true also for full QCD
in appropriate limits.
\end{abstract}

\end{frontmatter}


\section{Introduction}

In the high-energy regime of QCD, an interesting new phenomenon is expected
to show up: parton saturation \cite{GLR,MQ}.
Saturation
changes qualitatively the usual
equation for the evolution of scattering cross sections
with the energy of the reactions,
namely the so-called Balitsky-Fadin-Kuraev-Lipatov 
(BFKL) equation \cite{BFKL},
by
introducing nonlinearities.
In turn, due to these nonlinearities,
the intrinsic stochasticity
of partonic evolution may start to have 
a sizable effect on observables.

The basic nonlinear evolution
equation beyond the linear BFKL equation
is the Balitsky-Kovchegov
(BK) equation \cite{B,K1} (or, alternatively, the 
Balitsky-Jalilian Marian-Iancu-McLerran-Weigert-Leonidov-Kovner (B-JIMWLK) 
equations \cite{B,JIMWLK})
which however neglects the stochastic effects alluded to before, 
and is therefore
a kind of mean-field approximation.
While a formulation which would include all stochastic effects
has not been fully achieved yet
(the most recent advances may be found in Ref.~\cite{AKLP}),
it is believed that the complete 
evolution has a lot in common with
some reaction-diffusion processes described by equations
of the Fisher-Kolmogorov-Petrovsky-Piscounov (FKPP) type \cite{VSP}.
At the mean-field level,
this analogy is a formal identity between the FKPP equation
and the BK equation
in the so-called diffusive limit and assuming uniformity in
the transverse space (i.e. the gluon distribution
is assumed to evolve in the same way at all points
of the impact-parameter space) \cite{MP1}.
Beyond the mean-field approximation,
the conjecture made so far for the full problem
is that at any fixed position in transverse space,
the rapidity evolution of say the gluon content of a hadronic object
is like the time evolution of a one-dimensional reaction-diffusion process,
whose space variable would be the logarithm of the transverse
size (or momentum) of the gluons. 
Generally speaking, the dynamics of such systems is described
by equations equivalent to stochastic 
extensions to the
FKPP equation. (The first ideas on
how to go beyond the BK equation were presented in Ref.~\cite{MSh};
the equivalence with reaction-diffusion processes
was conjectured in Ref.~\cite{IMM};
deeper insight can be found in e.g. Ref.~\cite{IT}, and a review
in Ref.~\cite{MPhysRep}.)

While this analogy is useful to find asymptotic properties of observables
that involve one unique impact parameter,
so far little is known about the correlations
and the fluctuations
of the gluon distribution between 
different points in transverse
space, which
would show up in observables that probe several
points in
impact-parameter space simultaneously.

The correlations of the gluon number densities 
at different points in transverse space
were computed exactly
in Ref.~\cite{HM} in the context of 
the systematic approximation
to full QCD
provided by the color dipole model \cite{M1}, 
which is an accurate representation of the physics 
described by the BFKL equation but which
does not take into account saturation effects. 
In Ref.~\cite{IM}, a calculation was done in a theory with
full saturation based on a similarity
with the Liouville gravity. But that calculation was valid only
up to distances of the order of the inverse saturation momentum.
On the numerical side, on one hand, the BK equation was solved taking
into account the full impact-parameter dependence
\cite{GBS}, and on the other hand, the relevance of
the one-dimensional stochastic
FKPP equation at each fixed impact parameter
was tested in toy models for QCD evolution beyond the mean-field
approximation~\cite{MSS}.

In this paper, we would like to investigate how the saturation
scale varies in impact-parameter space
in the presence of both saturation and fluctuations.
Our method will consist in proposing 
a simple toy model which contains
the main physical features of QCD,
which may be implemented as a Monte-Carlo event generator and
for which analytical calculations will be possible.
In these respects, our approach follows the one developed
in Ref.~\cite{MSS}, but while the latter work was
purely numerical, our main results will consist in
analytical expressions of the correlation of
the saturation scale between two points in impact-parameter
space, as a function of the distance between the points
and as a function of the rapidity.

The model is introduced in the next section. We then
provide the derivation of the analytical expression
for the correlations. Finally, we check our calculations
against numerical simulations of different versions of
the model.


\section{Toy model}

\subsection{Construction}

The model that we introduce here is a simplified version of
the model proposed in Ref.~\cite{MSS}.

The starting point is the 
QCD color dipole model \cite{M1},
supplemented with some {ad hoc} saturation mechanism which limits
the number of dipoles in any given phase space cell.
The dipole model accurately represents the QCD evolution
in the high-energy regime and in the
limit of a large number of colors.
It provides an equation for the change of the density of
gluons (represented by a set of color dipoles
of different sizes and positions in the 
two-dimensional
plane transverse
to the flight axis of the hadron) inside a hadron,
when rapidity is increased. The basic process
from which the evolution
is built is the splitting of a dipole represented by
its two endpoints $(x_0,x_1)$
into two dipoles $(x_0,x_2)$ and $(x_1,x_2)$ with
the rate~\cite{M1}
\be
\frac{dP}{d(\bar\alpha Y)}=
\frac{x_{01}^2}{x_{02}^2x_{21}^2}d^2x_2.
\label{eq:dPdY}
\ee
As usually, $\bar\alpha=\alpha_s N_c/\pi$, where $\alpha_s$
is the strong coupling constant, $N_c$ the number of colors
and $Y$ is the rapidity.
The splitting of dipoles 
is a linear process, which generates the
BFKL equation when averages over dipole configurations
(``events'' in an experimental language) are taken.
The rapidity $Y$ is an effective evolution time.

When the rapidity becomes very high, then
gluons and thus dipoles may start to interact
among themselves, which induces nonlinear terms in
the evolution equations.
The effect of these interactions
is to tame the growth of the phase-space number density
of dipoles as soon as it reaches $N\sim 1/\alpha_s^2$,
which would otherwise be exponential with the rapidity.
The precise mechanism for these effects
is still not known in QCD, but the main observables
should be quite independent of these details.

With respect to QCD, we assume
the following simplifications:
{\em (i)} Dipoles evolve by giving birth
to one dipole of half size
(the left or the right half of the parent dipole),
or to one dipole of double size (in such a way
that the parent be the left or right half of its
offspring) at some
fixed rates,
{\em (ii)} dipoles do not disappear in the evolution, 
that is to say, the parent
dipoles are not removed,
{\em (iii)} the positions and dipole sizes are discrete,
and {\em (iv)} the configuration space of the dipoles is a
line instead of the full two-dimensional space.
We thus
give up two main properties of the QCD dipole model:
The collinear singularities, which cause
the dipole endpoints to emit an arbitrary number
of dipoles of arbitrarily
small sizes, and the continuous and two-dimensional nature
of the dipole sizes and positions.
The first simplification is the diffusion approximation,
which has been studied in the context of BFKL physics
(see e.g. Ref.~\cite{CC}),
but which was not assumed in Ref.~\cite{MSS}.
The second simplification was instead already assumed in~\cite{MSS}.
These model simplifications may introduce some artefacts, but
that we believe are under control, and many results
which we will obtain within such simple
models are likely
to apply to QCD since
they will not depend on the details.

\begin{figure}
\begin{center}
\epsfig{file=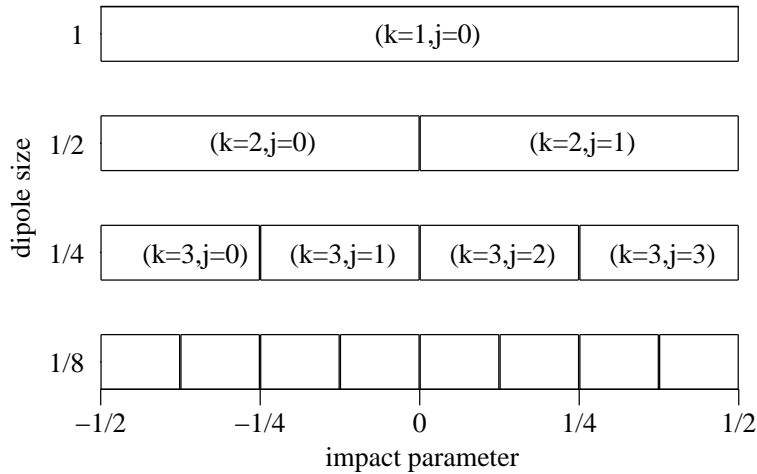,width=10cm}
\end{center}
\caption{\label{fig:sketch0}The hierarchical 
structure of the model.
Each box represents a bin which may contain 
up to $N$ dipoles of given sizes (vertical
axis) and positions in impact-parameter space (horizontal
axis). The conventional
numbering of the bins 
that we have chosen
is also shown for $k=1,2,3$.
}
\end{figure}

Let us now specify completely the model.
According to the evolution rules given above,
starting from a dipole of size 1,
the sizes of all dipoles present in the system
after evolution are powers of 2.
In practice, we shall only consider fractions of 1, i.e.
the sizes may be written as $2^{1-k}$, where $k\geq 1$.
For each value of $k$, there are $2^{k-1}$ possible
values of the position $b$ of the center of the dipoles: 
$b=-\frac12+2^{-k},-\frac12+3\times 2^{-k},\cdots,
\frac12-3\times 2^{-k},\frac12-2^{-k}$.
Let us number these bins
by the index $0\leq j \leq 2^{k-1}-1$ running
from the negative to the positive positions.
The model may be represented as 
a hierarchy of bins that contain
a discrete number of dipoles, see Fig.~\ref{fig:sketch0}.
Note that to any given impact parameter $b$ between 
$-\frac12$ and $\frac12$
corresponds one unique bin at each level of
size.
For example, at position $b=-\frac12$, one sees the bins
$(k=1,j=0)$, $(k=2,j=0)$, $(k=3,j=0)$ etc...
At position $-0.2$, one sees the bins
$(k=1,j=0)$, $(k=2,j=0)$, $(k=3,j=1)$ etc...
More generally, at position $-\frac12+\Delta b$, one sees 
$(k,[\Delta b\times 2^{k-1}])$, where
the square brackets represent the integer part.

During the rapidity 
(or time)
interval $dt$, a dipole
in the bin $(k,j)$
has a probability $\alpha dt$ to give
birth to a dipole in the bin $(k+1,2j)$,
$\alpha d t$ to give a dipole in the bin 
 $(k+1,2j+1)$, and $\beta dt/2$
to give a dipole
in the bin $(k-1,j/2)$ if
$j$ is even and $(k-1,(j-1)/2)$ if $j$ is odd.
Note that $dt$ may be infinitesimal (which is
generally speaking convenient
for analytical calculations), but
also finite (which is convenient
for numerical simulations).

As for the saturation mechanism, we
assume the simplest one: We veto splittings to
bins which already host the number $N$ of dipoles.

We can consider that the number density of ``gluons'' of
a given size
seen at one impact parameter
is proportional to the number of dipoles in the corresponding
bin $(k,j)$.
As rapidity is increased, the occupation of the
bins with low values of $k$ gets higher until the
number of objects they contain reaches $N$.
The subsequent filling of the bins indexed by larger
values of $k$ (smaller dipole sizes)
can be seen as the propagation of
traveling wave fronts at each impact parameter, 
with possibly complicated
relationships between them.
The (logarithm of the) saturation scale $X(b,t)$
at impact parameter $b$
is related to the position of the front
seen there at time $t$.
There are several equivalent ways to define
the position of the front.
It could be, for example,
the largest value of $k$ for which the number of
objects becomes some given
fraction of $N$. (Later, we will use a slightly different
definition).


\subsection{Basic features of the model}

Let us denote by $n_{(k,j)}(t)$ the number of dipoles
present in bin $(k,j)$ at time $t$.
Then, according to the rules given above, we can write 
the following stochastic evolution equation:
\begin{multline}
n_{(k,j)}(t+dt)=\min\bigg[N,n_{(k,j)}(t)
+\delta^\alpha_{(k-1,[j/2])}(t)\\
+\delta^{\beta/2}_{(k+1,2j)}(t)+\delta^{\beta/2}_{(k+1,2j+1)}(t)
\bigg],
\label{eq:defmodel1}
\end{multline}
where the $\delta^x_{(k,j)}$ are drawn according to the binomial
distribution
\be
\mbox{Proba}\left[\delta^x_{(k,j)}(t)=l\right]=
\begin{pmatrix}{n_{(k,j)}(t)}\\{l}\end{pmatrix}
(x dt)^l(1-x dt)^{n_{(k,j)}(t)-l}.
\label{eq:defmodel2}
\ee
This is a rather complicated equation which we do not
know how to solve except numerically. 

This model does not a priori look
like a stochastic FKPP model.
We may assume uniformity in impact parameter: This would
amount to imposing the same $\delta^\alpha$ and $\delta^{\beta/2}$
respectively
for all $j$ at any given $k$.
In this case, the model
would be projected to the FKPP class, but
by definition, this would wash out the
fluctuations between the different impact parameters.
This simplified model, 
that we call ``FIP'' (for ``Fixed Impact Parameter'',
since effectively, the model is completely
defined by a single impact parameter) 
in the terminology of Ref.~\cite{MSS}, is nevertheless
useful since it provides a benchmark
to evaluate how the fluctuations
between different impact parameters
may alter the FKPP picture.
In this paper, we will rely on 
(and check again in the case of our model)
the conclusion 
reached in Ref.~\cite{MSS}
that thanks to saturation,
locally at each impact parameter,
the full model is still well-described by
a one-dimensional FKPP equation,
and the fluctuations 
between different positions in impact-parameter space
do not qualitatively change the picture.

Let us first apply the well-known treatment of FKPP
equations to the FIP case. 
We know that the large-rapidity realizations
of the model are stochastic traveling waves, whose
main features can be determined from a simple analysis
of the linear part of the evolution equation.
In this model, only the number of dipoles $n_k$
in the bins
say $(k,0)$ (i.e. at impact parameter $-\frac12$) is relevant.
The evolution equation reads\footnote{%
We could also write $2\delta^{\beta/2}_{k+1}(t)$
instead of the last term in Eq.~(\ref{eq:evolutionFIP}).
(This may even be a more literal implementation
of the FIP approximation).
But this would not make a large difference,
which anyway, we would be unable to capture analytically.
}
\be
n_k(t+dt)=\min
\bigg[
N,n_k(t)+\delta^\alpha_{k-1}(t)+\delta^\beta_{k+1}(t)
\bigg].
\label{eq:evolutionFIP}
\ee
The mean-field (or Balitsky-Kovchegov) 
approximation to the evolution leads to the equation
\be
n_k(t+dt)=\min
\bigg[
N,n_k(t)+\alpha dt\, n_{k-1}(t)
+\beta dt\, n_{k+1}(t)
\bigg],
\ee
where the $n_k$ are now real functions of $k$.
The linearized equation (equivalent to the BFKL equation) 
is simply obtained by
discarding the ``$\min$'' in the previous equation:
\be
n_k(t+dt)=
n_k(t)+\alpha dt\, n_{k-1}(t)+\beta dt\, n_{k+1}(t).
\label{eq:BFKL}
\ee
From standard arguments, 
we know that for asymptotically large $t$ and $N$, the
velocity of the wave front, that is the time derivative
of the position $X(t)$ of the front, is given by
\cite{VSP,GLR,GBMS,MT}
\be
v_0=\frac{dX}{dt}=\chi^\prime(\gamma_0),
\label{eq:v0}
\ee 
where $\chi(\gamma)$ is the
eigenvalue of the kernel of the linearized evolution 
equation~(\ref{eq:BFKL}) corresponding to the 
eigenfunction~$e^{-\gamma k}$, namely
\be
\chi(\gamma)=\frac{1}{dt}\ln
\left(1+
\alpha dt\,e^\gamma+\beta dt\,e^{-\gamma}
\right),
\label{eq:eigenvalue}
\ee
and $\gamma_0$ minimizes $\chi(\gamma)/\gamma$.
We recall that $dt$ may be finite or infinitesimal,
in which case Eq.~(\ref{eq:eigenvalue}) is to be understood
as the derivative of $\ln$.
The shape of the front is a decreasing exponential
to the right of the saturation region,
\be
n_k(t)\sim N e^{-\gamma_0 (k-X(t))},
\label{eq:exposhape}
\ee
extending to $k\rightarrow\infty$ for large times.

The corrections due
to the discreteness of $n_k$ are known.
Taking the latter into account,
the 
front has now a finite extension, of the order of
\be
L_0=\frac{\ln N}{\gamma_0}.
\label{eq:L0}
\ee
Indeed, typically, it cannot extend further than the
point where $n_k(t)\sim 1$, and Eq.~(\ref{eq:L0})
then follows from Eq.~(\ref{eq:exposhape}).
Its mean velocity reads \cite{BD,MSh}
\be
v_\text{BD}=\chi^\prime(\gamma_0)
-\frac{\pi^2\chi^{\prime\prime}(\gamma_0)}
{2\gamma_0 L_0^2}.
\label{eq:vBD}
\ee
This velocity was obtained in a still deterministic (mean-field)
approximation with appropriate cutoffs to mimic saturation 
and discreteness.
Taking furthermore fluctuations into account, the velocity becomes
\cite{BDMM}
\be
v=\chi^\prime(\gamma_0)
-\frac{\pi^2\chi^{\prime\prime}(\gamma_0)}
{2\gamma_0\left(L_0+\frac{3}{\gamma_0}\ln L_0\right)^2},
\label{eq:vBDMM}
\ee
and realization-to-realization fluctuations are
characterized by the following cumulants of the 
position of the front:
\be
\frac{[\mbox{$n-$th cumulant}]}{t}
=\frac{\pi^2\chi^{\prime\prime}(\gamma_0)
n!\zeta(n)}{\gamma_0^{n+1}L^3},
\label{eq:cumBDMM}
\ee
where a priori the model leads to $L=L_0$, but empirically,
a better fit
to the results of the
numerical simulations is obtained by adding a 
subleading correction of
the form
$L=L_0+3\ln L_0/\gamma_0+\mbox{const}$.
In particular, the diffusion constant of the front,
that is to say the slope of the $t$-dependence of the
second-order cumulant reads
\be
D=\frac{\pi^4\chi^{\prime\prime}(\gamma_0)}{3\gamma_0^3L^3}.
\label{eq:D}
\ee
These expressions are valid in the limit of large $L\sim L_0$,
i.e. for exponentially large values of $N$.

There exists 
a subclass of these models that may be reduced exactly to
a collection of FKPP models.
Let us set $\beta=0$, that is to say, authorize only splittings
to smaller-size dipoles.
Consequently, there may not be any influence of the bins at any
size
level $k$ on the content of the bins of level less than $k$
(i.e. of larger sizes). Then, at each impact parameter,
one has a FIP model, i.e. a model
of the one-dimensional FKPP type.
However, the relationship between the
different copies of FIP models
is not trivial, since part of the evolution
is common between different impact parameters.
Even for $\beta\ne 0$, we believe that this minimal model,
consisting in considering two one-dimensional systems
appropriately correlated, is a good approximation
to the full model.

In our investigations, we will have in mind the latter
class of models, and we will check numerically 
that for more general
models for which $\beta\ne 0$, the results that we shall
obtain are not significantly altered.


\section{Correlations in the hierarchical model}

Our aim 
is to study the correlations between the point
at position $b=-\frac12$ in transverse space
(left edge of the system, 
see Fig.~\ref{fig:sketch0})
and the one at position $b=-\frac12+\Delta b$ with
$0\leq \Delta b<1$.
We calculate the
average of the squared difference of the
positions of the front
between these
points, which is formally related to the
two-point correlation function
of (the logarithm of) the saturation scales,
and which we deem a good
estimator of the spatial fluctuations of the saturation
scale. In the hierarchical model,
all bins with index $k$ less than 
or equal to
$k_{\Delta b}\equiv 1+[-\log_2\Delta b]$ 
(the notation ``$[\cdots]$''
stands for the integer part) and $j=0$
overlap both impact parameters, and thus the
dipoles of size larger than $2^{-k_{\Delta b}}$
seen at these points are exactly the same.
For $k>k_{\Delta b}$ instead, the bins seen
at the two points are distinct and nonoverlaping.
So in particular,
in our model with $\beta=0$, as soon as
the position of the front
at one point or at the other is
larger than $k_{\Delta b}$, that is to say, as soon
as there are of the order of $N$ 
dipoles in the bin $(k_{\Delta b},j=0)$, then the evolutions are
completely uncorrelated
at the two points in the corresponding bins.
(We expect that for finite $\beta$ of order 1,
the discussion would not be qualitatively changed.)
This matches to the picture that we may infer for the
QCD dipole model: The dipoles at two
positions in impact-parameter space
separated by a distance larger than the typical saturation
scales in that region evolve (almost)
independently towards larger
rapidities.
Note that choosing pairs of points
around impact parameter 0, one with positive impact parameter
and another one with negative impact parameter,
 would not satisfy this property,
due to the rigidity of the sizes and positions of the dipoles.
Indeed, these two points would decorrelate very soon in 
the evolution since their common ancestors necessarily
sit in the bin $(k=1,j=0)$, 
see Fig.~\ref{fig:sketch0}.

As a consequence of these features of 
QCD reproduced in the toy model,
studying two-point
correlations between points in 
impact-parameter space
as a function of their distance $\Delta b$ and of the 
time (=rapidity) $t$
is equivalent to
studying the time dependence of the correlations of
the saturation scales of two realizations of the
model whose evolutions
are identical until the tip of the front
reaches $k_{\Delta b}$. On the average,
it takes a time $t_{\Delta b}=(k_{\Delta b}-1)/v$, $v$ being
the mean velocity of the individual
fronts, 
for the front whose tip is at $k_{\Delta b}=1$ at the beginning of the
evolution to have its tip at $k_{\Delta b}$.
Then the bins such that $k>k_{\Delta b}$
evolve independently between the two realizations
over the remaining 
time interval 
\be
\Delta t=t-t_{\Delta b},
\ \ \mbox{with}\ \ 
t_{\Delta b}=\frac{\left[-\log_2\Delta b\right]}{v}.
\label{eq:ttb}
\ee
Note that
this is very close to assuming that the realizations are identical
for $t\leq t_{\Delta b}$ and completely
uncorrelated for $t>t_{\Delta b}$.

From this discussion, we see that
the basic input of our calculation will
be the mechanism for the propagation of a FKPP front.
We will review it in the next subsection, then we will
proceed to the formulation of the calculation of
the correlations.


\subsection{Short review of the mecanism for stochastic 
front propagation}

In this section, we review the 
recent progress in the understanding
of the mecanism for front propagation \cite{BD,MSh,BDMM},
which has eventually
led to the formulation of a model, 
and from which Eqs.~(\ref{eq:vBDMM}),~(\ref{eq:cumBDMM}) 
follow \cite{BDMM}.

Instead of solving the nonlinear problem, it was
proposed \cite{BD,MSh} to replace the nonlinearities
by absorptive boundaries, and to treat the evolution
equation as a linear (branching diffusion) equation 
between these boundaries.
There are two types of nonlinearities: {\em (i)} The
saturation condition that 
keeps the number of particles in each bin less than or equal to
$N$ and
{\em (ii)} the very discreteness of this number of particles,
whose effect mainly shows up in the region where $n\sim 1$.
The position of the boundaries is adjusted in such a way that
the distance between them be the size of 
the front~(\ref{eq:L0}),
namely $L_0=\ln N/\gamma_0$, and that the large-time solution of
the diffusion equation be stationary.
This procedure leads to the expression of the front velocity
for large $N$ given in Eq.~(\ref{eq:vBD}), 
but it does not predict the fluctuations of
the position of the front.

In order to incorporate the latter,
we add to this picture the possibility that
there be one (or a few) particles 
randomly
sent ahead of the
tip of the front~\cite{BDMM}, at a distance $\delta$. 
We attribute to this event 
a probability per unit time
\be
p(\delta)d\delta=C_1 e^{-\gamma_0\delta}\Theta(\delta)d\delta,
\label{eq:p}
\ee
in the continuation of the shape 
of the front~(\ref{eq:exposhape})
solution in the large-$t$ and large-$N$ limit. 
($C_1$ is some constant, undetermined at this stage).
In Ref.~\cite{BDMM}, we treated such fluctuations
as localized extra weights of appropriate ``mass'' a unit
or so to the left of the right boundary.
We then computed the effect of this weight
on the position of the front at large time, and found a forward
shift of the position equal to
\be
R(\delta)=\frac{1}{\gamma_0}\ln
\left(
1+C_2 \frac{e^{\gamma_0\delta}}{L_0^3}
\right).
\label{eq:R0}
\ee
$C_2$ is another constant.
At large times after a fluctuation has occurred, the
front relaxes to its mean-field shape.
We assumed that the relevant fluctuations are
rare enough in such a way that the front has time to
completely
relax between two fluctuations,
which turns out to be true for $L_0\gg 1$.

From the two quantitative
elements~(\ref{eq:p}) and~(\ref{eq:R0}), together
with the solution of the stationary diffusion 
problem between the boundaries~(\ref{eq:vBD}), we may
write down an effective theory for stochastic front propagation,
with however two unknown parameters, namely $C_1$ and $C_2$.
While we were not able to determine $C_1$ and $C_2$
separately, it is the product $C_1C_2$ that appears in all
cumulants of the position of the front,
and in particular in the correction to the velocity
induced by fluctuations.
Assuming that the velocity of the front taking into account 
the fluctuations
is the velocity of a mean-field front whose size is extended by
$3 \ln L_0/\gamma_0$ with respect to $L_0$
(i.e. with the substitution $L_0\rightarrow L_0+3\ln L_0/\gamma_0$), 
we got a determination of the product $C_1C_2$:
\be
C_1C_2=\pi^2\chi^{\prime\prime}(\gamma_0).
\label{eq:C1C2}
\ee
The effective theory for front propagation then leads to
Eqs.~(\ref{eq:vBDMM}) and~(\ref{eq:cumBDMM}).
As we will see below, the only new ingredient that will
be needed is the time dependence of the shift of the
front $R$, whose large-time asymptotics is Eq.~(\ref{eq:R0}).


\subsection{Formulation of the calculation of correlations}

In line with the above discussion, we wish to compute the
correlations of the position of two fronts whose evolutions are
identical for $t\leq t_{\Delta b}$ and uncorrelated for $t>t_{\Delta b}$.
Note that strictly speaking, we would need to
keep the content of all bins $k\leq k_{\Delta b}$ identical 
between the two realizations at all times, even
after time $t_{\Delta b}$.
But these two formulations give quantitatively similar
results.

Let us introduce $X(t_0,t)$ the position of the front at time $t$ in the
frame in which $X(t_0,t_0)=0$.
We focus on what happens slightly before the initial time $t_0$.
On one hand, $X(t_0-dt_0,t)=X(t_0,t)+v_\text{BD}dt_0$
if no fluctuation has occurred between times $t_0-dt_0$ and $t_0$,
on the other hand,
$X(t_0-dt_0,t)=X(t_0,t)+v_\text{BD}dt_0+R(t-t_0,\delta)$ if a 
fluctuation has
occurred at a position $\delta$ ahead of the front
(which happens with probability $p(\delta)d\delta\,dt_0$).
It is straightforward to
 write an equation for the generating function of the
cumulants of $X$:
\be
-\frac{d}{dt_0}\ln\left\langle e^{\lambda X(t_0,t)}\right\rangle
=\lambda v_\text{BD}+\int d\delta\,p(\delta)
\left(
e^{\lambda R(t-t_0,\delta)}-1
\right).
\ee
One now considers two such independent 
fronts and add up the generating functions. One gets
\begin{multline}
-\frac{d}{dt_0}\ln\left(
\left\langle
e^{\lambda X_1(t_0,t)}
\right\rangle
\left\langle
e^{-\lambda X_2(t_0,t)}
\right\rangle
\right)\\
=\int d\delta\,p(\delta)
\left(e^{\lambda R(t-t_0,\delta)}+e^{-\lambda R(t-t_0,\delta)}-2
\right).
\end{multline}
Expanding for $\lambda$ close to 0,
the coefficients of the second power of $\lambda$ 
obey the equation
\be
\frac{d}{dt}\left\langle (X_1-X_2)^2\right\rangle
=2\int d\delta \,p(\delta) R^2(t-t_0,\delta),
\label{eq:corr}
\ee
where we have used the fact that $X_1$ and $X_2$ are independent
random variables for $t>t_0$, and we have
traded $t_0$ for $t$ in the derivative, taking advantage
of the fact that both $X_1-X_2$ and $R$ only depend
on $t-t_0$.
In practice, $t_0$ will be equal to $t_{\Delta b}$, the time
at which the tip of the single front reaches $k_{\Delta b}$.
From Eq.~(\ref{eq:ttb}), this time is $[-\log_2\Delta b]/v$.

We see that the basic ingredient is 
the time evolution of the shift of
the front due to a forward fluctuation.
This shift was given in Eq.~(\ref{eq:R0})
in the limit of large times. We have to repeat the steps
that led to Eq.~(\ref{eq:R0}) keeping however track
of the full time dependence.


\subsection{Effect of a fluctuation on the position of the front}

The problem amounts to solving a diffusion equation
between two fixed absorptive boundaries, with various
initial conditions.
We shall discuss the 
scaled dipole number $u(t,k)=n_k(t)/N$,
and for the simplicity of the formalism, 
consider that it is a function of a real variable $k$.

Let us write the general branching diffusion equation:
\be
\partial_t u(t,k)=\chi(-\partial_k)u(t,k),
\label{eq:equ}
\ee
where $\chi$ is an appropriate kernel
that encodes the linear evolution of the
dipoles. (The
operators that appear here are denoted as differential
operators, but they
could also be finite differences
as in Eq.~(\ref{eq:BFKL}).
Analytical calculations
are usually easier with differential operators).
In the case of QCD, $\chi(-\partial_k)$ would be
the BFKL kernel (that can easily be deduced
from Eq.~(\ref{eq:dPdY})), $t\sim\bar\alpha Y$,
$k\sim\ln 1/r^2$ (where $r$ is the size of the dipoles),
and $u$ would be the scattering amplitude.
Further, we define $\gamma_0$ to be the eigenvalue of $\chi$
that satisfies $\chi(\gamma_0)=\gamma_0\chi^\prime(\gamma_0)$.
Following Ref.~\cite{BDMM}, we write the ansatz
\be
u(t,k)=e^{-\gamma_0(k-X(t))}
L\psi\left(\frac{2\chi^{\prime\prime}(\gamma_0)t}{L^2},
\frac{k-X(t)}{L}\right).
\label{eq:ansatz}
\ee
$L$ is the size of the front, which is essentially $L_0=\ln N/\gamma_0$
for large $N$.
When $\chi$ is expanded to second order around
the eigenvalue $\gamma_0$,
then $\psi$ obeys the partial differential equation
\be
\partial_y\psi=\frac{1}{4}\partial_\rho^2\psi
+\frac{\gamma_0 L^2}{2\chi^{\prime\prime}(\gamma_0)}
(\chi^\prime(\gamma_0)-X^\prime(t))\psi,
\label{eq:diffusion0}
\ee
where we have defined
\be
y=\frac{2\chi^{\prime\prime}(\gamma_0)t}{L^2}\ \ 
\text{and}\ \ 
\rho=\frac{k-X(t)}{L}.
\label{eq:changevar}
\ee
We have only kept the dominant terms
for large $L$.
We see that $\chi^\prime(\gamma_0)-X^\prime(t)$ has to scale
like $1/L^2$ for all terms of this equation to be relevant.
The coefficient must be chosen in such a way that
in the large-$y$ limit, there is a nontrivial stationary solution.
We shall use the already known result~(\ref{eq:vBD}) to write
\be
X^\prime(t)=\chi^\prime(\gamma_0)-\frac{\pi^2\chi^{\prime\prime}(\gamma_0)}
{2\gamma_0 L^2}+o(1/L^2)
\label{eq:Xprime}
\ee
and check a posteriori that it is the correct expression.
Equation~(\ref{eq:diffusion0}) then becomes
\be
\partial_y\psi=\frac{1}{4}\partial_\rho^2\psi
+\frac{\pi^2}{4}
\psi,
\label{eq:diffusion}
\ee
up to higher-order terms when $L$ is large.

We shall admit that the saturation of the number of dipoles
and the stochasticity may be appropriately implemented,
within well-controlled approximations, by
respectively an absorptive boundary at
$\rho=0$ and another one at $\rho=1$ (which corresponds to a distance
$L$ between the boundaries in $k$-coordinates, i.e. to the natural
width of the stationary front which travels at the 
velocity $v_\text{BD}$).
The boundary conditions formally read
\be
\psi(y,\rho=0)=0\ \ \mbox{and}\ \ \psi(y,\rho=1)=0.
\label{eq:bc}
\ee

Let us now discuss the initial condition.
The fluctuations that will generate the front-shifts $R$
responsible for the corrections to $v_\text{BD}$
are dipoles sent at a distance $\delta$ ahead of the tip of the
steady front.
Formally, in order to describe such a fluctuation occurring
at time $t=0$, we would write
$u(0,k)=\delta(k-X(0)-L-\delta)/N$, 
i.e. from Eq.~(\ref{eq:ansatz}),
$\psi(0,\rho)=e^{\gamma_0\delta}
\delta(\rho-1-\delta/L)/L^2$.
But since there is an absorptive boundary 
at $\rho=1$,
we had better
move the fluctuation to the left
of the right boundary.
This should not qualitatively change the problem
as long as $L\gg 1$.
Thus we write
\be
\psi(y=0,\rho)=\delta(\rho-1+\bar a)\frac{e^{\gamma_0\delta}}{L^2},
\label{eq:ic}
\ee
where $\bar a$ is a constant of order $1/L$, and therefore $\bar a\ll 1$.

The solution of Eq.~(\ref{eq:diffusion}) with the conditions~(\ref{eq:bc})
and~(\ref{eq:ic}) reads
\be
\psi_\delta(y,\rho)=
\frac{2e^{\gamma_0\delta}}{L^2}
\sum_{n=1}^\infty
(-1)^{n+1}
\sin\pi n \bar a\,
\sin\pi n \rho\,
e^{-\frac{\pi^2 (n^2-1) y}{4}}.
\label{eq:psi}
\ee
The subscript $\delta$ recalls that this solution
was obtained
starting with an initial condition consisting
in one dipole at a distance $\delta$ ahead of
the deterministic front.

We first discuss the stationary solution.
We see that for large $y$, 
the higher harmonics are suppressed
exponentially with respect to the fundamental mode $n=1$,
which gives the following contribution:
\be
\psi_{\delta_0}(y,\rho)=
\frac{2e^{\gamma_0\delta_0}}{L^2}
\sin \pi \bar a
\sin \pi \rho
\underset{\bar a\ll 1}{\simeq}
\frac{2\pi\bar a e^{\gamma_0\delta_0}}{L^2}\sin\pi\rho.
\label{eq:psi0}
\ee
Thanks to the choice~(\ref{eq:Xprime}) for $X'(t)$,
this solution has no $y$ dependence, and leads
to a stationary $u$ in the frame of the front.
The expression~(\ref{eq:psi0}) is independent of the initial
condition except for the overall normalization.
The value of $\delta_0$,
which characterizes the initial condition,
will be adjusted later.
Undoing the changes of variables which trade
$u$ for $\psi$, $k$ for $\rho$ and $t$ for $y$
(Eq.~(\ref{eq:ansatz})),
the stationary solution $u_{\delta_0}$ reads
\be
u_{\delta_0}(t,k)=
e^{-\gamma_0(k-X(t))}\frac{2\pi \bar a e^{\gamma_0\delta_0}}{L^2}
\left[
L\sin\frac{\pi(k-X(t))}{L}
\right].
\label{eq:u0}
\ee
We require that $u_0(t,k)\sim 1$ for $k=X(t)+aL$,
where $aL$ is a constant of order~1.
This condition is satisfied if we set 
$\delta_0\sim 3\ln L/\gamma_0$.
Indeed, with this choice,
\be
u_{\delta_0}(t,X(t)+aL)\simeq 2\pi^2\bar a a L^2 e^{-\gamma_0 a L}.
\label{eq:pointnorm}
\ee
Since $\bar a L$ and $a L$ are constants,
the right-hand side is just a number of order 1.

We now add a fluctuation to the stationary front, 
that is to say,
an extra particle at a position $\delta$ ahead of the tip
of the steady front.
The solution
of the diffusion equation
is the superposition of the large-time stationary solution
$u_{\delta_0}$ given by Eq.~(\ref{eq:u0})
with~$\delta_0=3\ln L/\gamma_0$,
and of the solution $u_\delta$ of the diffusion equation
with the generic initial condition
characterized by $\delta$
(see Eq.~(\ref{eq:psi})), up to a multiplicative
constant $C_2$ of order 1 that we do not control
in this calculation, since it certainly 
depends on the detailed shape
of the fluctuations. 
We write
\be
\begin{split}
u(t,k)&=u_{\delta_0}(t,k)+C_2 u_\delta(t,k)\\
&=e^{-\gamma_0(k-X(t))} L \left[\psi_{\delta_0}(y,\rho)+C_2 \psi_\delta(y,\rho)
\right]
\end{split}
\label{eq:u}
\ee
up to the replacement of the variables
by their expressions~(\ref{eq:changevar}).
\begin{figure}
\begin{center}
\epsfig{file=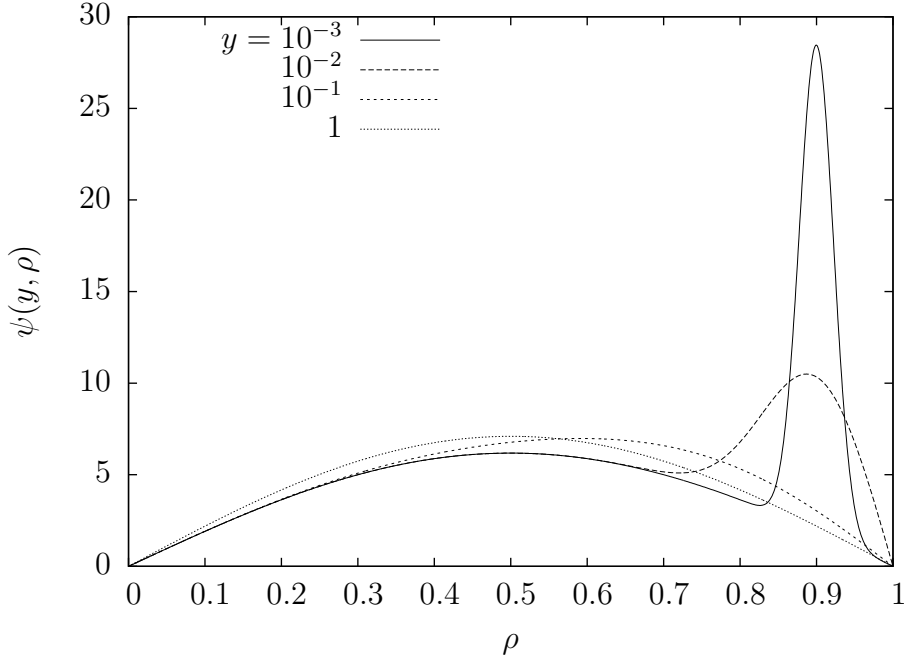,width=13cm}
\end{center}
\caption{\label{fig:psi}$\psi_{\delta_0}(y,\rho)+\psi_\delta(y,\rho)$
[see Eqs.~(\ref{eq:psi0}), (\ref{eq:psi})]
for different values of 
the reduced time variable $y$
after a fluctuation of size $\delta=5$ has occurred at $y=0$.
The size of the front is $L=10$, and $\bar a=0.1$.
We see how the fluctuation, initially localized at the tip of the front,
gets smeared uniformly over the width of the front as $y$ gets large.
Eventually, a small forward shift $X\rightarrow X+R$ 
would be needed in order to
absorb it and recover the stationary front.
}
\end{figure}
The presence of the
second term alters the shape of the front
(the front eventually relaxes back to the sine shape
in Eq.~(\ref{eq:u0})),
see Fig.~\ref{fig:psi}. %
But of course,
we want to keep the normalization condition
for $u$, namely for some appropriate value of $k$,
$u$ is required to equate Eq.~(\ref{eq:pointnorm})
at all $t$. This is possible by shifting
the value of $k$ at which we 
enforce the normalization condition 
from $k=X(t)+aL$ to say $k=X(t)+aL+R(t,\delta)$.
This is equivalent to shifting the
position of the front
$X(t)\rightarrow X(t)+R(t,\delta)$. 
Equation~(\ref{eq:u}) then leads to
\be
u(t,X(t)+aL+R(r,\delta))
=2\pi^2 \bar a a L^2
e^{-\gamma_0 a L-\gamma_0 R(t,\delta)} \left[1+
C_2\frac{\psi_\delta(y,a)}{2\pi^2 \bar a a\, L}
\right].
\label{eq:ufin}
\ee
Equating the right-hand sides of
Eq.~(\ref{eq:ufin}) and Eq.~(\ref{eq:pointnorm}),
we get
\be
R(t,\delta)=\frac{1}{\gamma_0}
\ln
\left[
1+{C_2} 
\frac{\psi_\delta\left(\frac{2\chi^{\prime\prime}(\gamma_0)t}{L^2},
a\right)}
{2\pi^2\bar a a\, L}
\right],
\label{eq:R}
\ee
where only the lowest orders in $\bar a$, $a$ 
in the expansion of $\psi_\delta$
must be kept.
With the help of Eq.~(\ref{eq:psi}), it is then
straightforward to arrive at an explicit 
expression of $R$.

Note that 
by performing the shift $X\rightarrow X+R$,
we have actually
added a $t$-dependent term to $X^\prime(t)$ in
Eq.~(\ref{eq:Xprime}), which was initially assumed to be a constant.
However, since the time scale for the variations
of $R$ is $L^2$, $R^\prime$
is suppressed
for large $L$, and thus 
Eq.~(\ref{eq:diffusion})
is not modified within our approximations.

It is interesting to note that $\psi_\delta$
is related to some Jacobi $\vartheta$ function \cite{AS}.
Since
\be
\vartheta_4(z|q)=1+2\sum_{n=1}^{\infty}(-1)^n\cos(2nz)q^{n^2},
\ee
we may rewrite Eq.~(\ref{eq:psi}) as
\be
\psi_\delta(y,\rho)=\frac{e^{\gamma_0\delta}}{2 L^2}
\frac{1}{q}
\bigg[
\vartheta_4\left(\frac{\pi(\bar a+\rho)}{2}\bigg|q\right)
-\vartheta_4\left(\frac{\pi(\bar a-\rho)}{2}\bigg|q\right)
\bigg].
\ee
The notation
\be
q\equiv e^{-\frac{\pi^2 y}{4}}
=e^{-\frac{\pi^2 \chi^{\prime\prime}(\gamma_0)t}{2L^2}}
\label{eq:q}
\ee
has been introduced.
Using Eq.~(\ref{eq:R}) and performing the appropriate expansion
for small $\bar a$ and $a$, we arrive at
an expression for $R(t,\delta)$ 
in terms of the $\vartheta_4$-function
which is particularly
compact:
\be
\boxed{
R(t,\delta)=\frac{1}{\gamma_0}\ln\left[
1-C_2\frac{e^{\gamma_0\delta}}{2 L^3}
\partial_q\vartheta_4(0|q)
\right],
}
\label{eq:Rtheta}
\ee
with
\be
-\partial_q\vartheta_4(0|q)=2\sum_{n=1}^{+\infty}
(-1)^{n+1}n^2 q^{n^2-1}.
\label{eq:theta4def1}
\ee
It is actually quite natural that the Jacobi theta functions appear,
since the latter are defined as solutions of the one-dimensional 
heat equation
with periodic boundary conditions.

We turn to the analysis of the obtained result.
First, for large $y$, 
only the fundamental mode contributes to $\psi_\delta$, and
it is clear that Eq.~(\ref{eq:R}) reduces to Eq.~(\ref{eq:R0}).
Looking back at Eq.~(\ref{eq:psi}), 
we see that higher
harmonics would give a series of exponentially
 decreasing corrections.
But at a finite time,
a large number of modes have to
be taken into account, typically all modes such that
$n\leq (L/\pi) \sqrt{2/\chi^{\prime\prime}(\gamma_0)t}$. 
A few low-lying modes are not enough
to describe the small-time behavior. Instead,
it is a saddle point
(in an appropriate integral reformulation)
that dominates the sum~(\ref{eq:psi}).
In this regime, it would be
useful to find a way to write the series of harmonics
such that at asymptotically large $y$, 
only the first term contributes instead of the whole
series. This is actually possible using the Poisson
summation formula
\be
\sum_{n=-\infty}^{+\infty}f(n)=\sum_{k=-\infty}^{+\infty}
\int dx\,f(x)e^{-2i\pi k x}.
\label{eq:poisson}
\ee
In order to get $R(t,\delta)$, we need the value 
of $\psi_\delta$ at $\rho=a$.
Hence we choose
\be
f(x)=-\frac{e^{\gamma_0\delta}}{L^2}
\sin\pi x \bar a
\sin\pi x a 
\,
q^{x^2-1}
e^{i\pi x}.
\ee
We then perform the integral
over $x$ in the r.h.s. of Eq.~(\ref{eq:poisson}).
Introducing $\gamma_+=\bar a+a$ and $\gamma_-=\bar a-a$,
we get the following expression for $\psi_\delta$:
\begin{multline}
\psi_\delta(y,a)=\frac{e^{\gamma_0\delta}}{4 L^2}
\frac{1}{q}\sqrt{-\frac{\pi}{\ln q}}
\sum_{k=-\infty}^{+\infty}
\bigg(
e^{\frac{(2k-1+\gamma_+)^2\pi^2}{4\ln q}}
+e^{\frac{(2k-1-\gamma_+)^2\pi^2}{4\ln q}}\\
-e^{\frac{(2k-1+\gamma_-)^2\pi^2}{4\ln q}}
-e^{\frac{(2k-1-\gamma_-)^2\pi^2}{4\ln q}}
\bigg).
\end{multline}
Since we eventually want to apply Eq.~(\ref{eq:R})
in order to get an expression of the shift of the front,
we expand the latter formula
for $\bar a,a\ll 1$.
The leading order reads
\be
\psi_\delta(y,a)=\frac{\sqrt{\pi}}{2}
\frac{e^{\gamma_0\delta}}{q(-\ln q)^{5/2}}
\frac{\pi^2 \bar a a}{L^2}
\sum_{k=1}^{+\infty}
\left[
\pi^2(2k-1)^2+2\ln q
\right]
e^{\frac{(2k-1)^2\pi^2}{4\ln q}}.
\ee
The shift of the
front due to a fluctuation
is obtained 
from $\psi_\delta$
with the help of Eq.~(\ref{eq:R}):
\begin{multline}
R(t,\delta)=\frac{1}{\gamma_0}
\ln\bigg\{
1+C_2
\frac{\sqrt{2}L^2 e^{\gamma_0\delta}}{(\pi\chi^{\prime\prime}(\gamma_0)t)^{5/2}}
e^{\frac{\pi^2\chi^{\prime\prime}(\gamma_0)t}{2L^2}}\\
\times\sum_{k=1}^{+\infty}
\left[(2k-1)^2-\frac{\chi^{\prime\prime}(\gamma_0)t}{L^2}
\right]
e^{-\frac{(2k-1)^2L^2}{2\chi^{\prime\prime}(\gamma_0)t}}
\bigg\}.
\label{eq:Rcomplet}
\end{multline}
$q$ is the function of $t$ given by Eq.~(\ref{eq:q}).
This formula is extremely useful, since the series
indexed by $k$ converges fast.
Even for moderately large values of $t$,
a few terms accurately describe the whole function.
This is actually the best formula for numerical
evaluations of $R$.

We shall now examine the limit of small $t$ ($y\ll 1$).
Then only the term $k=1$ has to be 
kept.
The expression for $R$ boils down to
\be
R(t,\delta)=\frac{1}{\gamma_0}
\ln\bigg(
1+C_2\frac{\sqrt{2}e^{\gamma_0\delta}L^2}
{(\pi\chi^{\prime\prime}(\gamma_0))^{5/2}}
\frac{e^{-\frac{L^2}
{2\chi^{\prime\prime}(\gamma_0)t}}}{t^{5/2}}
\bigg).
\ee

As a final remark, let us note that 
the Poisson summation~(\ref{eq:poisson})
 that we have used to rewrite the
series of harmonics corresponds to a Jacobi identity
for the $\vartheta$ functions \cite{AS}.
Equation~(\ref{eq:Rcomplet}) results from
Eq.~(\ref{eq:Rtheta}) with the replacement
\be
-\partial_q\vartheta_4(0|q)=\frac{\sqrt{\pi}}{2}
\frac{1}{q(-\ln q)^{5/2}}
\sum_{k=1}^{+\infty}\left(\pi^2(2k-1)^2+2\ln q\right)
e^{\frac{(2k-1)^2\pi^2}{4\ln q}}.
\label{eq:theta4def2}
\ee


\subsection{Analytical expression for the correlations}

With the elements presented in the previous sections,
we can write the expression for 
$\sigma_{12}^2\equiv\langle (X_1-X_2)^2\rangle$.
It is enough to insert the expression for the probability 
of fluctuations
(Eq.~(\ref{eq:p}))
and for the time-dependent shift (Eq.~(\ref{eq:Rtheta}))
into Eq.~(\ref{eq:corr}):
\be
\frac{d\sigma_{12}^2}{dt}=\frac{2C_1}{\gamma_0^2}
\int_0^{+\infty}d\delta\,e^{-\gamma_0\delta}
\ln^2\left[
1-C_2\frac{e^{\gamma_0\delta}}{2 L^3}
\partial_q\vartheta_4(0|q)
\right],
\label{eq:calsigma2}
\ee
where for $\partial_q\vartheta_4(0|q)$ we use 
either one of the equivalent
expressions~(\ref{eq:theta4def1}),~(\ref{eq:theta4def2})
according to the limit that we want to investigate.
We now have to fix the value of $L$.
In Ref.~\cite{BDMM}, $L$ was taken to be a constant.
(The phenomenological
model predicted $L=L_0\equiv \ln N/\gamma_0$,
but empirically, we saw that it 
was better to add a subdominant correction,
namely $L=L_0+\frac{3}{\gamma_0}\ln L_0+\mbox{const}$.)
In this case, a change of variable can be made
in the integrand. All the parameters may be factored out,
leaving us with a simple numerical integral to perform:
\be
\int_0^{+\infty}\frac{dx}{x^2}\ln^2(1+x)=2\zeta(2)=\frac{\pi^2}{3}.
\ee
Thus
\be
\frac{d\sigma_{12}^2}{dt}=\frac{\pi^2C_1C_2}{3\gamma_0^3L^3}
\left[-\partial_q\vartheta_4(0|q)\right].
\ee
Replacing the product of the unknown constants by 
Eq.~(\ref{eq:C1C2}) and $q$ by Eq.~(\ref{eq:q}) and integrating
over the time variable between 0 and $\Delta t=t-t_{\Delta b}$, 
we arrive at a parameter-free expression for
$\sigma_{12}^2$ as a function of $\Delta t$, namely
\be
\boxed{
\sigma_{12}^2=\frac{2\pi^2}{3\gamma_0^3L}
\int_{e^{-\frac{\pi^2\chi^{\prime\prime}(\gamma_0)\Delta t}{2L^2}}}^1
\frac{dq}{q}\left[-\partial_q\vartheta_4(0|q)\right].
}
\label{eq:sigma122f}
\ee

We now investigate the two interesting limits,
i.e. $\Delta t\gg L^2$ and $\Delta t\ll L^2$.
For large $\Delta t$, 
the integral is dominated by the region
$q\rightarrow 0$, thus
$-\partial_q\vartheta_4(0|q)$
may be replaced by its value at $q=0$ ($-\partial_q\vartheta_4(0|0)=2$).
Performing the remaining integration, we get
\be
\sigma_{12}^2\underset{\Delta t\gg L^2}{\sim}
\frac{2\pi^4\chi^{\prime\prime}(\gamma_0)}
{3\gamma_0^3 L^3}\Delta t,
\ee
which is twice the second-order cumulant of the position
of the FIP front, see Eq.~(\ref{eq:D}).
For small $\Delta t$ instead, say $L\ll \Delta t\ll L^2$, 
we use the expansion
of $\partial_q\vartheta_4(0|q)$
for $q\rightarrow 1$, i.e. the first term in Eq.~(\ref{eq:theta4def2}),
which reads
\be
\partial_q\vartheta_4(0|q)=
-\frac{\sqrt{\pi}}{2}\frac{\pi^2+2\ln q}{q(-\ln q)^{5/2}}
e^{\frac{\pi^2}{4\ln q}}.
\ee
Equation~(\ref{eq:sigma122f})
boils down to the following expression:
\be
\sigma_{12}^2\underset{\Delta t\ll L^2}{\sim}
\frac{4}{3\gamma_0^3}
\sqrt{
\frac{2\pi^3}{\chi^{\prime\prime}(\gamma_0)\Delta t}
}
\exp\left(-\frac{L^2}{2\chi^{\prime\prime}(\gamma_0)\Delta t}\right).
\label{eq:smalltasymptotics}
\ee

So far, we have chosen 
the size of the front
$L$ constant, of the order of $L_0$.
Another possible model for $L$
would be to promote it to a function of $\delta$
at the level of Eq.~(\ref{eq:calsigma2}), namely
\be
L=L_0+\delta+\mbox{const},
\ee
where the constant has to be determined empirically.
This choice takes maybe into account more accurately
the extension of the front by $\delta$ that generates
the fluctuations.
The $\delta$-integral cannot be performed
analytically in Eq.~(\ref{eq:calsigma2}) except in 
some limits, so
a priori,
there is no simpler expression than Eq.~(\ref{eq:calsigma2}). 
Thus
we need to know the values of $C_1$ and $C_2$
individually. 
We can consider that
$C_1=\gamma_0$ is the natural normalization
of the probability distribution $p(\delta)$. Then, we
must set $C_2=\pi^2\chi^{\prime\prime}(\gamma_0)/\gamma_0$
in order to satisfy Eq.~(\ref{eq:C1C2}).

The above-mentioned
two models,
in which $L$ is either constant or $\delta$-dependent,
differ by subleading terms in the large-$L$
limit.
Since the values of $\delta$ which dominate the $\delta$-integral
in Eq.~(\ref{eq:calsigma2})
are of order $\frac{3}{\gamma_0}\ln L_0$, 
like the first correction to $L_0$ in the case of constant $L$,
the models are not
expected to differ significantly.
We will check this statement numerically.


\subsection{Scaling}

Looking back at Eq.~(\ref{eq:sigma122f}),
 we see that $\sigma_{12}^2$ has a nice
scaling property.
Indeed, we may rewrite the latter equation as
\be
\sigma_{12}^2=\frac{D}{\gamma_0(v_0-v)}
\int_{e^{-\gamma_0(v_0-v)\Delta t}}^1\frac{dq}{q}
\left[-\partial_q\vartheta_4(0|q)\right]
\ee
in terms of the properties of a single front (its velocity $v$
and the diffusion constant $D$ whose analytical expressions
were given in Eq.~(\ref{eq:vBDMM}) and~(\ref{eq:D})), 
where $v_0$ can be read in Eq.~(\ref{eq:v0}).
In particular, we have the following scaling:
\be
\boxed{
\frac{\sigma_{12}^2}{D\Delta t}=\mbox{function}[(v_0-v)\Delta t].
}
\label{eq:scaling}
\ee
From Eq.~(\ref{eq:smalltasymptotics}), 
we see that the function in the right-hand side
is exponentially damped when
its argument is smaller than 1, i.e. parametrically
for $\Delta t\ll L^2$.

Once one knows the characteristics of the traveling waves
in the FIP model (i.e. $v$ and $D$), this
scaling of the correlations is a pure prediction.
Thus it will be interesting to check it in the
numerical calculations.


\subsection{Limits on the validity of the calculations}

Let us try and
evaluate the limits on the validity of our calculations.
The latter were essentially based on the assumption 
that the eigenvalue $\gamma=\gamma_0$
of the kernel $\chi$ dominates.
While this statement is clearly true 
at large times, when the traveling-wave
front is well formed, (see e.g. Ref.~\cite{MT}), 
it must break down at early
times right after a fluctuation has occurred:
Indeed, a fluctuation
has an initial 
shape that is far from the one of the asymptotic front, 
see Fig.~\ref{fig:psi}.

We wish to estimate the order of magnitude of the dispersion
of the relevant eigenvalues
about $\gamma_0$. To this aim, 
neglecting for the moment the boundary conditions
and the prefactors,
we write the solution of Eq.~(\ref{eq:equ}) as
\be
u(\Delta t,k)\sim \int d\gamma\, e^{-\gamma k+\chi(\gamma)\Delta t}.
\ee
The interesting values of $k$ are the ones around the 
position of the
wave front,
therefore we write
$k=v_0 \Delta t+\delta k$, where $\delta k$ is of the order
of the size $L$ of the front.
Expanding
$\chi(\gamma)$ about $\gamma_0$, we write
\be
u(\Delta t,v_0\Delta t+\delta k)\sim 
e^{-\gamma_0\times\delta k}
\int d(\delta\gamma)\,
e^{-\delta\gamma\times\delta k +\frac12 \chi^{\prime\prime}(\gamma_0)
(\delta\gamma)^2\Delta t+\cdots},
\label{eq:uapprox}
\ee
where $\delta\gamma=\gamma-\gamma_0$.
It is clear from this equation that the relevant values
of $\delta\gamma$ are of the order of 
$\delta k/(\chi^{\prime\prime}(\gamma_0)\Delta t)$.
Since the order of magnitude of $\delta k$ is the size $L$ of
the front,
we would a priori conclude that the dispersion of $\gamma$ 
around $\gamma_0$ is small
and hence that the calculation is valid as soon as $\Delta t\gg L$.

However, we have also expanded $\chi(\gamma)$ to second order.
This means that for a generic kernel $\chi$, we have neglected terms
of the form
$\frac16\chi^{(3)}(\gamma_0)(\delta\gamma)^3\Delta t\sim L^3/(\Delta t)^2$ 
(which would fit in the
dots in Eq.~(\ref{eq:uapprox})).
The expansion is a good approximation if the latter 
term is small, i.e. if
\be
\Delta t\gg L^{3/2}.
\ee


\subsection{Back to impact-parameter space}

So far, we have been working with the minimal model,
consisting in two realizations of the FIP model
which evolve in the same way until their 
common tip reaches $k_{\Delta b}$,
and which decorrelate for $k>k_{\Delta b}$.
The only relevant parameter 
which determined the decorrelation of the positions 
of the fronts
of the realizations
was the time $\Delta t=t-t_{\Delta b}$ 
after the tip had reached $k_{\Delta b}$.
We now wish to discuss the transcription of the
obtained results
to impact-parameter space, which was our initial problem.

To this aim, we will of course 
make use of Eq.~(\ref{eq:ttb})
to express $t_{\Delta b}$ with the help of the mean front
velocity $v$. 
But we also need a length scale to which 
the distance in impact-parameter space
$\Delta b$
may be compared. The natural length
is the dipole size at the position 
of the front, namely
\be
l_s(t)=2^{-X(t)}=l_s(t_{\Delta b})2^{-v\Delta t}.
\label{eq:ls}
\ee
On the other hand, according to Eq.~(\ref{eq:ttb})
and disregarding the integer part operator,
$-\log_2\Delta b=k_{\Delta b}$ and the tip of the front
$k_{\Delta b}$ is ahead of the bulk $X(t_{\Delta b})$
by $L$: $k_{\Delta b}=X(t_{\Delta b})+L$. Using the previous equation,
we may now express $\Delta t$ as a function of $\Delta b$ and of
the length scale $l_s(t)$:
\be
\Delta t=\frac{1}{v}\left[L+\log_2\frac{\Delta b}{l_s(t)}\right].
\ee
The scaling~(\ref{eq:scaling})
reads
\be
\sigma_{12}^2\sim
\frac{L+\log_2\frac{\Delta b}{l_s(t)}}{L^3}\times
\mbox{function}
\left[\frac{L+\log_2\frac{\Delta b}{l_s(t)}}{L^2}\right].
\ee
This formula, together with the behavior of
the scaling function (see Eq.~(\ref{eq:smalltasymptotics})),
shows that there is little $b$-dependence
until $\log_2 (\Delta b/l_s(t))\sim L^2$, that is to say,
until $\Delta b\sim l_s(t) e^{\text{const}\times L^2}$.
In other terms, the size $\Delta b$ of the
domain around impact parameter $b$ in which the 
fluctuations in the
position of the fronts
are negligible is, in notations more familiar to QCD experts,
\be
\boxed{
\Delta b\sim \frac{e^{\text{const}\times \ln^2(1/\alpha_s^2)}}{Q_s(b)},
}
\ee
where $Q_s(b)$ is the usual saturation momentum at impact
parameter $b$.
Note that since the fronts are statistically independent
as soon as $\Delta b\times Q_s(b)>1$, this result may seem a bit surprising:
It says that the effective correlation length between different
points in impact-parameter space is much larger than $1/Q_s(b)$
in the parametrical limit of small $\alpha_s$.
This is the main qualitative result of this paper.


\section{Numerical simulations}

In this section, we confront our analytical calculations
to numerical simulations of the toy model.
First, we consider the full model and
test the validity of the
assumption that the minimal model
is a good approximation to the full model
also for $\beta\sim 1$, i.e. when splittings
to larger-size dipoles are authorized.
Second,
we compare the minimal model to the analytical results
for the fluctuations between different positions in
impact-parameter space
(given essentially by
Eqs.~(\ref{eq:calsigma2}),(\ref{eq:sigma122f})).

\subsection{Full model}

The model defined by Eqs.~(\ref{eq:defmodel1}) and~(\ref{eq:defmodel2}) 
is straightforward to implement numerically
in the form of a Monte-Carlo event generator.
The simplest is to store
the number of dipoles in each bin
in an array whose index $i$ is related to $k$ and $j$
through $i=2^{k-1}+j$. 
The splitting dynamics relates bin $i$ to $2i$ (down left),
$2i+1$ (down right) and $[i/2]$ (up; the square brackets stand 
once again for
the integer part).

We have to deal
with an array whose size grows exponentially with time.
It is thus very difficult to pick large values of $t$,
and thus also large values of $N$. Indeed, the relevant time
scale grows with $N$ like $\ln^2 N$, and consequently
the minimum number of
entries in the array 
one wants to consider
grows like $e^{\ln^2 N}$.
In practice, we limit ourselves to $t\leq 4$ and $N\leq 100$.
As for the time step $dt$, the most convenient is to take it
small but finite. We set $dt=10^{-2}$.

\begin{figure*}
\begin{center}
\epsfig{file=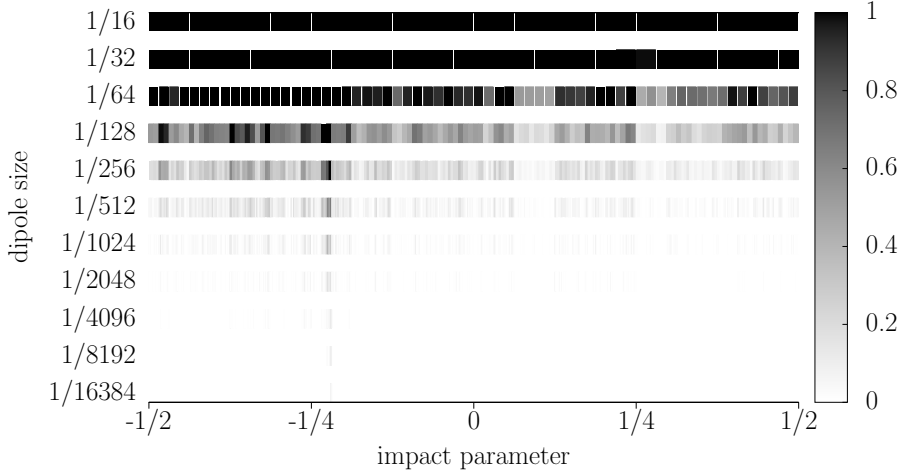,width=13cm}
\end{center}
\caption{\label{fig:sketch1}One event of the full model
with $\alpha=\beta=1$, $N=100$ and $t-t_{\Delta b}=4$. 
Only the bins $k\geq 5$ are
represented. (The bins for $k<5$ all contain $N$ dipoles.)
The number of dipoles in each bin is proportional to the blackness
which is displayed.
We see that in the transition region close to blackness, 
nearby bins are often of similar grey levels, which
illustrates the statement that the density of gluons varies significantly
only over scales which are larger than the relevant
length scale $l_s(t)$ (see Eq.~(\ref{eq:ls})).
}
\end{figure*}

We start with one particle and evolve it for a few hundred units of time
using the FIP version of the
model. We obtain a traveling wave
front, whose tip we eventually label $k=1$. (The complete front
sits in the bins $k\leq 1$).
From the initial condition built in this way, we evolve
all bins for which $k<1$ using the FIP model, and
all bins for
$k\geq 1$ using the full model.
One event in shown in Fig.~\ref{fig:sketch1}.
Although $N$ and $t$ are small in this calculation, 
we see that the regions 
in impact-parameter
space which have similar numbers of dipoles are larger
than the local length scale $l_s(t)$ (see Eq.~(\ref{eq:ls})).

After the evolution times $t=3$ and $t=4$ respectively, 
we measure the position of the front at various impact
parameters on a uniform tight grid
ranging from $-\frac12$ to
$+\frac12$.
We use the following definition 
of the position of the front:
\be
X(\Delta b,t)=k_0
+\sum_{k=k_0+1}^{+\infty}\frac{n_{(k,[\Delta b\times 2^{k-1}])}(t)}{N}
\ee
where $k_0$ is the largest $k$ for which $n_{(k,[\Delta b\times 2^{k-1}])}=N$.
Note that in principle, we could have chosen $X(\Delta b,t)=k_0$.
In practice however, because of the discreteness of $k$
in our model, this choice
would introduce artefacts which we do not expect in real QCD.

We compute the squared difference of the front positions
between the impact parameters $-\frac12$ and $-\frac12+\Delta b$,
and average over events.
We plot the result as a function of $t+\log_2\Delta b/v$,
where $v$ is the average front velocity measured
at impact parameter $-\frac12$.

We compare the results to the correlations
obtained in the minimal model, i.e.
when we consider two independent 
realizations of an initial front.
We do not attempt to compare to our analytical formulas
since the values of $N$ that we are able to reach
are too small for the approximations
that we had to assume to be relevant.

\begin{figure}
\begin{center}
\epsfig{file=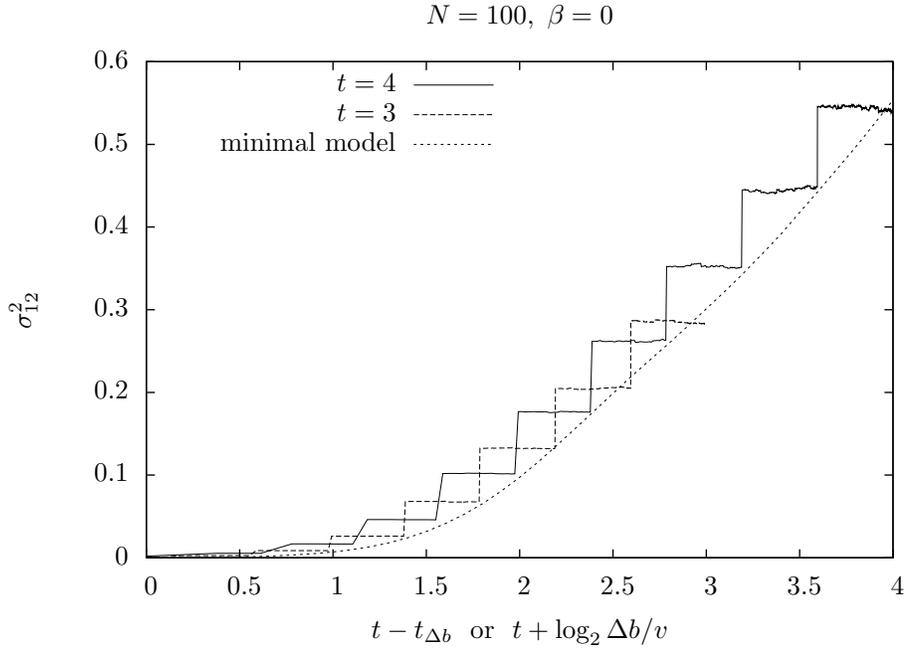,width=13cm}
\end{center}
\caption{\label{fig:plot_100_0}
$\sigma^2_{12}=\langle (X_1-X_2)^2\rangle$ as a function
of $\Delta t=t-t_{\Delta b}$ in the full model with $\alpha=1$ and $\beta=0$ 
(lines with steps; one corresponds to an evolution time $t=3$, the other
one to $t=4$)
and in the minimal model.
In the full model, $t_{\Delta b}=[-\log_2\Delta b]/v$, where $v$ is the measured
velocity of the front at impact parameter $-\frac12$.
In the FIP model, $t_{\Delta b}$ is a fixed time, and corresponds to
the time at which 
the tip of the front reaches $k_{\Delta b}$, the bin after which 
two uncorrelated evolutions take place.
}
\end{figure}

\begin{figure}
\begin{center}
\epsfig{file=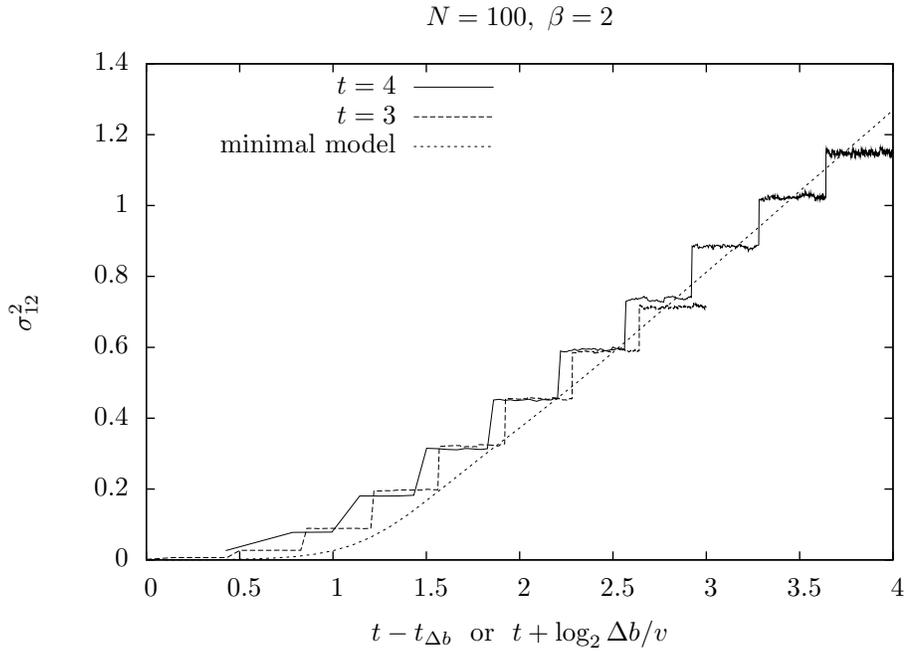,width=13cm}
\end{center}
\caption{\label{fig:plot_100_2}
The same as in Fig.~\ref{fig:plot_100_0} but for $\beta=2$.
}
\end{figure}

The corresponding plot is displayed in Fig.~\ref{fig:plot_100_0}
for $N=100$, $\alpha=1$, $\beta=0$, and in  Fig.~\ref{fig:plot_100_2}
with the same parameters except $\beta=2$. 
First, we see that in the full model,
the graph of $\sigma_{12}^2$ exhibits steps, i.e. 
$\sigma_{12}^2$ is constant by parts. 
This is related
to the hierarchical structure of the model:
The correlations between $b=-\frac12$ and
any of the points at $b\geq 0$ are identical;
The same is true for $-0.25\leq b<0$, $-0.375\leq b<-0.25$ etc...
The logarithmic $b$-scale on the $t$-axis makes the widths of the
steps all equal.
Next, we see that for small $t-t_{\Delta b}$ 
(i.e. impact parameters close
to $-\frac12$)
there are very little fluctuations in the front positions.

Finally, we see that for $\beta=0$, as anticipated, the full model
and the minimal ones coincide almost perfectly (Fig.~\ref{fig:plot_100_0}).
For $\beta=2$, 
i.e. when splittings towards larger dipole sizes are switched on
and therefore new correlations appear beyond
the ones taken into account in the minimal model,
there are some quantitative differences for large
$t$ (Fig.~\ref{fig:plot_100_2}). 
But we see that using the minimal model instead of the
full model that keeps all impact parameters is
a good approximation.
This corroborates the conclusions of the work in Ref.~\cite{MSS}.

\subsection{Minimal model}

\begin{figure}
\begin{center}
\epsfig{file=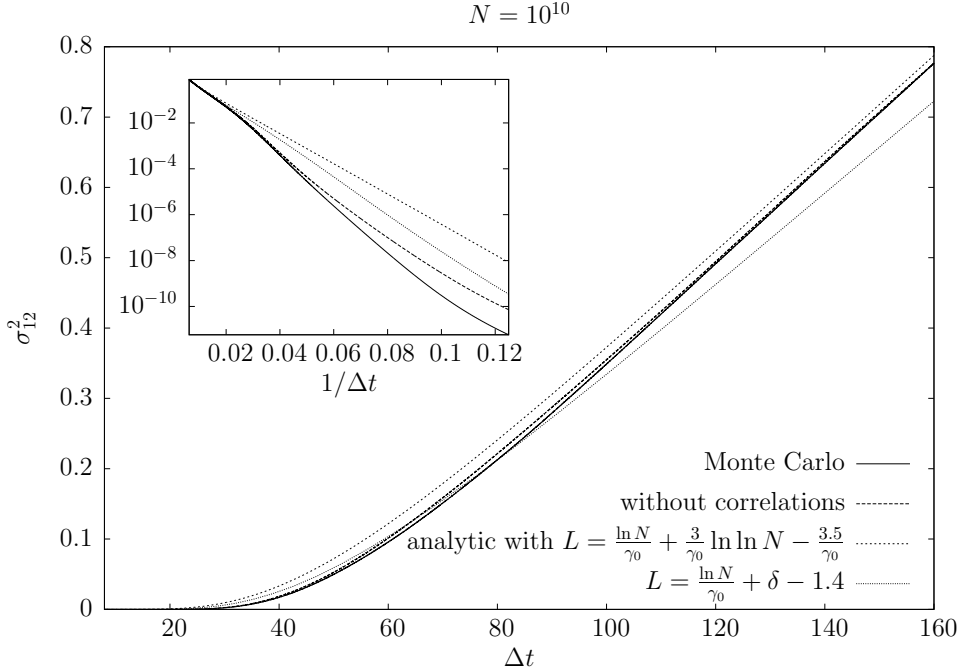,width=13cm}
\end{center}
\caption{\label{fig:plot10}
$\sigma_{12}^2$ as a function of $\Delta t=t-t_{\Delta b}$
in the minimal model with $\beta=0$
for $N=10^{10}$.
We display the results obtained within
the model in which the realizations
decorrelate in the bins $k>k_{\Delta b}$
(labelled ``Monte Carlo''), and within the model
in which the decorrelation is complete
after time $t_{\Delta b}$ (labelled ``without correlations'').
The theoretical curves use Eq.~(\ref{eq:sigma122f})
with the two possible choices for the front size $L$.
{\it Inset:} The same, as a function
of $1/\Delta t$ in order to highlight the
small-$\Delta t$ region where, as expected, important 
differences
appear between the models.
}
\end{figure}

\begin{figure}
\begin{center}
\epsfig{file=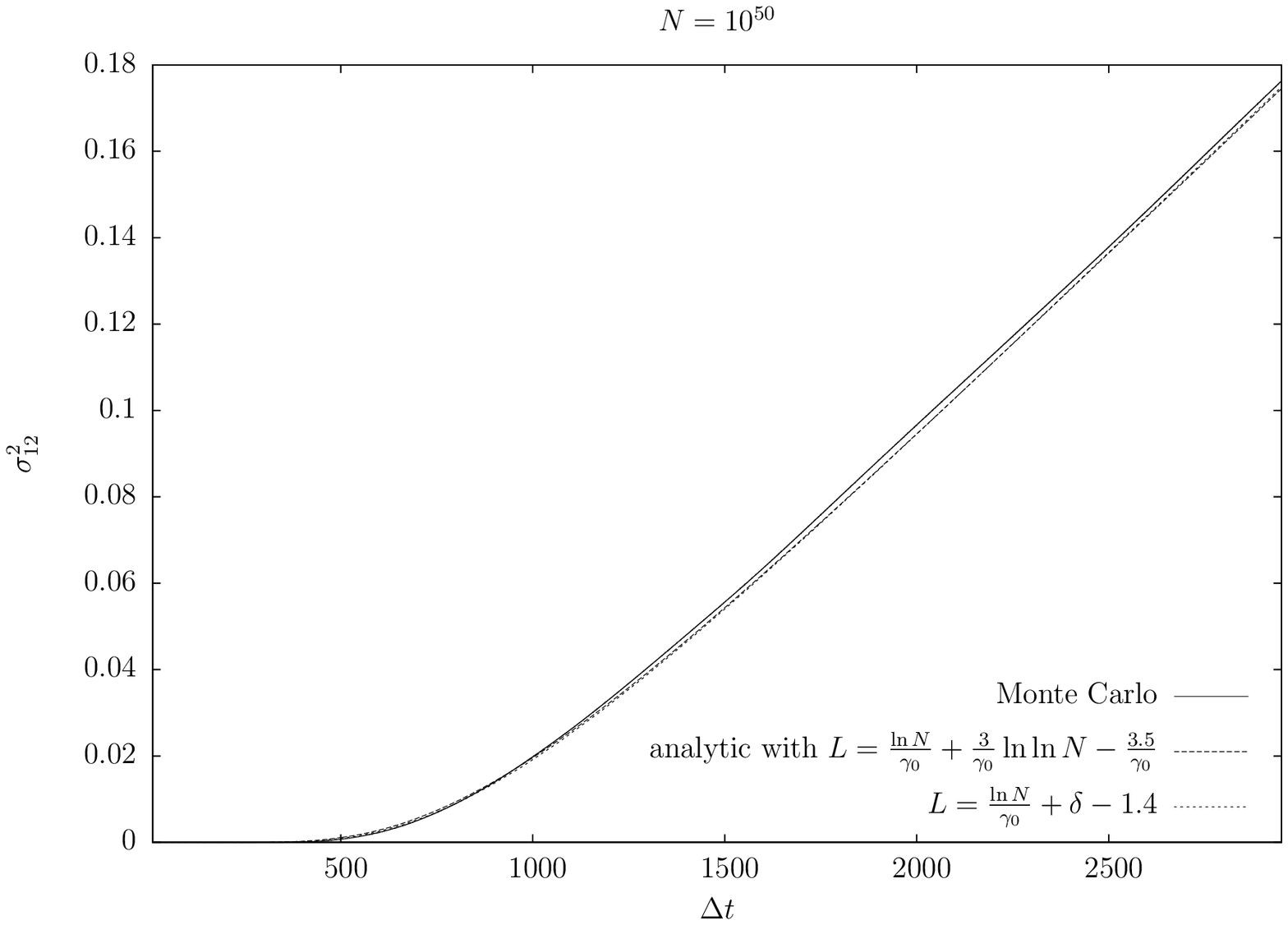,width=13cm}
\end{center}
\caption{\label{fig:plot50}
The same as in Fig.~\ref{fig:plot10}, for $N=10^{50}$.
All curves coincide almost perfectly.
}
\end{figure}

We now set $\beta=0$, in which case, as discussed
earlier and as checked numerically, 
the model exactly reduces to a collection
of one-dimensional FKPP-like models.
Hence, in order to compute two-point correlation functions,
it is enough to evolve two realizations of the corresponding
FIP model with the constraint
that all bins with $k\leq k_{\Delta b}$ be identical between the
two realizations, and the bins $k>k_{\Delta b}$ be completely
independent.
Alternatively, we could also generate one single realization
and evolve it for $t_{\Delta b}$ time steps, 
replicate it at time $t_{\Delta b}$,
and then evolve the two replicas completely independently of each
other. The difference between these two possible implementations
of the minimal model cannot be accounted for
in our analytical calculations,
thus the differences that we shall find 
numerically
will give an indication of 
the model uncertainty.
This time, our aim is essentially
to check our analytical formulas,
thus we will pick very large values of $N$,
even if they appear to be 
unphysical in the QCD context since
they would correspond to exponentially small values of
the strong coupling constant $\alpha_s$.

The parameters of the model are obtained from Eq.~(\ref{eq:eigenvalue})
with $\alpha=1$, $\beta=0$ and $dt=10^{-2}$:
\be
\gamma_0=1.0136\cdots\ ,\ \ v_0=2.6817\cdots\ ,\ \ 
\chi^{\prime\prime}(\gamma_0)=2.6098\cdots
\ee
These values are close to $1$, $e$ and $e$ respectively,
which would be the correct parameters if $dt$ were infinitesimal,
in which case $\chi(\gamma)=e^\gamma$ 
(see Eq.~(\ref{eq:eigenvalue})).

The numerical results are shown 
in Fig.~\ref{fig:plot10}
for $N=10^{10}$
with the two versions of the model
(we generated about $10^5$ realizations),
and compared with
the analytical predictions.
We test the two possible choices for the size $L$ of the
front: Either $L$ is a constant, which from our
previous experience with FKPP traveling waves \cite{BDMM},
we set to
\be
L=\frac{1}{\gamma_0}\ln N+\frac{3}{\gamma_0}\ln\ln N-\frac{3.5}{\gamma_0}
\ee
or it is $\delta$-dependent, namely
\be
L=\frac{1}{\gamma_0}{\ln N}+\delta-1.4.
\ee
The numerical
constants, which are not determined in our theory,
were chosen empirically so that they properly
describe all numerical data for $N\geq 10^{10}$.
In the first case, Eq.~(\ref{eq:sigma122f}) 
is used. In the second case,
Eq.~(\ref{eq:calsigma2}) 
is integrated numerically over
$t$ and $\delta$.
We see that the agreement between the numerical calculation
and the analytical predictions
is good, except maybe for very small values of $\Delta t$
where the calculations are not expected to be accurate.
Indeed, for the same values of $\Delta t$, we also 
see in Fig.~\ref{fig:plot10} a sizable discreapancy between
the two versions of the minimal model.
The calculations for $N=10^{50}$ are shown in Fig.~\ref{fig:plot50}.
The numerical results and the theoretical 
expectations (Eq.~(\ref{eq:sigma122f}))
coincide almost perfectly.

\begin{figure}
\begin{center}
\epsfig{file=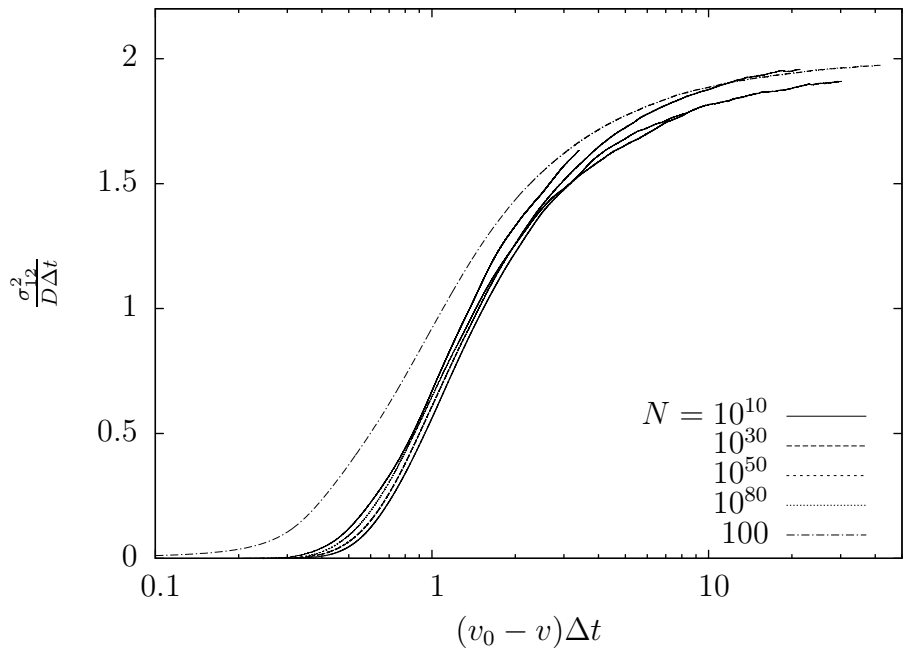,width=13cm}
\end{center}
\caption{\label{fig:scaling}
Numerical check of the scaling~(\ref{eq:scaling}).
The curves for the different values of $N$
are very close together for $N\geq 10^{10}$, 
but the scaling seems to break down for low values of $N$
(see the curve for $N=100$), as expected.
}
\end{figure}

Finally, we check that the scaling in Eq.~(\ref{eq:scaling})
is well reproduced by the numerical data.
The Monte-Carlo simulations are shown in Fig.~\ref{fig:scaling},
plotted in the appropriate scaling variables.
The diffusion constant of a single wave front $D$ as well as
the velocity $v$ are measured from the same data.
We see that all curves nicely superimpose for $N\geq 10^{10}$ (we
show data for values of $N$ as large as $10^{80}$),
while there are clear deviations for smaller $N$
(see the curve for $N=100$), as expected.

\section{Conclusion}

In this paper, we have built a model that
possesses the main features of the QCD dipole model
including the dynamics in impact-parameter space,
and which is furthermore very easy to implement numerically.
We have obtained analytical expressions for the
fluctuations of the (logarithm of the)
saturation scale
from one position in impact-parameter space to another one nearby,
which gives an indication on the homogeneity of
the gluon number density.

Since our analytical calculations are only based on some
rather general properties of the model,
they should go over to full QCD
after appropriate replacement of the parameters, hopefully giving the
correct small-$\alpha_s$ (large $N$) asymptotics.

We have found that the saturation scale varies quite
slowly. In the usual notations of QCD, 
if at position $b$ in impact-parameter space
the local saturation scale is $Q_s(b)$,
then the saturation scale is uniform over
a region of size $e^{\text{const}\times\ln^2(1/\alpha_s^2)}/Q_s(b)$
around that position.

Our calculations suffer the usual limitations
in this kind of models: Analytical expressions
are derived for exponentially large $N$, that is to say,
exponentially small values of the strong coupling constant
$\alpha_s$. Although they often may be successfully extrapolated
down to $N\sim 100$ ($\alpha_s\sim 0.1$), 
this is at the cost of tuning constants,
and so far, we have not found a systematic procedure to compute
finite-$N$ corrections.

The next step would probably be a numerical calculation
in the case of full QCD, 
using, as proposed earlier \cite{M2005},
a combination
of a Monte Carlo implementation of
the color-dipole model in the low-density regime 
(such an implementation is already available; 
see Ref.~\cite{S1,S_MC,MS} 
and Ref.~\cite{AGL} for more recent work)
and of
a numerical solution of the BK equation at the transition
to saturation \cite{GBS}. 
This looks very challenging,
but maybe the nice smoothness of the saturation scale
in impact-parameter space that we have found in
this paper will help.

Finally, it would be interesting to find an observable
which would directly be sensitive to the gluon density
at two points in impact-parameter space simultaneously.


\section*{Acknowledgments}

This work was supported in part by 
the Department of Energy (USA),
and in part by the 
Agence Nationale pour la Recherche (France),
contract ANR-06-JCJC-0084-02.


\end{document}